\documentclass [prb,aps,twocolumn,amsmath,amssymb]{revtex4}
\usepackage{latexsym,amsmath,graphics,graphicx}
\newcommand{\otop}[1]{\overset{\circ}{#1}\,\!}

\begin{document}
%\maketitle
\title{Surface state scattering by adatoms on noble metals}

\author{Samir Lounis}\email{s.lounis@fz-juelich.de}
\author{Phivos Mavropoulos}
\author{Peter H.~Dederichs}
\author{Stefan Bl\"ugel}
\affiliation{Institut f\"ur
Festk\"orperforschung, Forschungszentrum J\"ulich, D-52425 J\"ulich,
Germany}

\date{\today}

\begin{abstract}
When surface state electrons scatter at perturbations, such as magnetic
or nonmagnetic adatoms or clusters on surfaces, an electronic resonance, 
localized at the adatom site, can develop below the 
bottom of the surface state band for both spin channels.
In the case of adatoms, these states have been found very recently in
scanning tunneling spectroscopy experiments\cite{limot,olsson} 
for the Cu(111) and Ag(111) surfaces. Motivated 
by these experiments, we carried out 
a systematic theoretical investigation of the electronic structure of these
surface states in the presence of magnetic and non-magnetic atoms on 
Cu(111). We found that Ca and all 3$d$ adatoms lead to a split-off state 
at the bottom of the surface band which is, however, not seen for the 
$sp$ elements Ga and Ge. The situation is completely reversed if the 
impurities are embedded in the surface: Ga and Ge are able to produce a 
split-off state whereas the 3$d$ impurities do not. The resonance 
arises from 
the $s$-state of the 
impurities and is explained in terms of strength and interaction nature 
(attraction or repulsion) of the perturbing potential.

\end{abstract}
\maketitle
\section{INTRODUCTION}
At crystal surfaces the symmetry is lowered: The three dimensional periodicity 
in the bulk is lowered to the two dimensional periodicity at the surface. 
This leads to the occurrence of two-dimensional surface states \cite{tamm,shockley}, 
which are spatially confined to the surface, since their wave functions decay rapidly 
into the crystal and are strongly damped in the vacuum. Surface states can exist 
only in regions of the two-dimensional Brillouin zone, where bulk Bloch states 
are not allowed. 
They are characterized by a two-dimensional Bloch vector $\vec {k}_{||}$ in this surface 
Brillouin zone, which describes the propagation in the surface plane. 
A projection of the 
bulk band structure to the surface plane can result in ${k}_{||}$-regions, where 
bulk states are 
forbidden. In these gaps of the projected bulk band structure surface states can occur 
provided that their energy is lower than the work function. These two 
conditions guarantee that the wave functions of the surface states 
decay exponentially both into the crystal and 
into the vacuum region. 

Recently a strong interest arose concerning an important physical effect 
associated with the interaction of a two dimensional 
surface state with the states of an adatom. 
It was shown that for Cu adatoms on Cu(111) a bound state splits off from the bottom 
of the Cu(111) surface state\cite{limot,olsson}. 
This effect was basically predicted by Simon \cite{simon} 
who stated that in two-dimensional free space any 
attractive potential has a bound state. 
Gauyacq {\it et al.} \cite{gauyacq} suggested that an adatom--induced localization 
of the surface state may be observed in STS as a peak 
appearing just below the surface band edge when a Cs adatom is deposited on Cu(111).

Using low--temperature scanning tunneling spectroscopy (STS), 
Limot {\it et al.}~\cite{limot} investigated silver and cobalt 
adatoms on Ag(111) as well as copper and cobalt adatoms on Cu(111). The 
bound state appears both for magnetic and 
nonmagnetic adatoms. Moreover, using a Newns--Anderson model the 
authors explained the results as arising from the coupling of the 
adatom's orbital (which was supposed to be the $s$--orbital) 
with the surface-state electrons, and being broadened by the interaction with bulk 
electrons of the same energy. On the other hand, 
Olsson {\it et al.} \cite{olsson} used the same type of experiment 
and performed pseudopotential calculations for single 
Cu adatoms on Cu(111). The calculated local density of states (LDOS) exhibits 
several adatom--induced peaks. Two of them are assigned to resonances deriving 
from the $ d_{z^2}$ atomic orbital and $sp_z$ 
hybrid orbitals. The third one corresponds to a localization of the surface state at 
the adatom without a specified orbital origin.
In fact, these adatom-induced peaks appearing at the bottom of 
the surface state were already observed by an other 
experiment \cite{madhavan}, again without a clear 
assignment of their origin. Davis {\it et al.}\cite{davis} have 
observed similar localized states 
in STS measurements for Cr atoms in the surface layer of Fe(001) surface. 
Ab-initio calculations of Papanikolaou {\it et al.} \cite{papanikolaou2} confirmed 
this and showed that similar localized states occur for many other 
impurities in the Fe(001) surface.

In this work we study the origin and the condition of existence for such 
an impurity--induced split-off state on the Cu(111) surface. For this purpose 
we have performed ab--initio 
calculations using the Korringa-Kohn-Rostoker (KKR) Green function method for 
impurities on 
the Cu(111) surface. We consider single impurities of the 3{\it d} 
and 4{\it sp} elements as adatoms on the Cu(111) 
surface and as impurities in the first layer. We 
find that split-off states can appear both for adatoms on the surface as well 
as for substitutional impurities in the surface. In the case of magnetic impurities 
these states always appear in both spin channels and show a very small spin splitting.

\section{Computational Aspects}

In the present work we use the 
full--potential KKR Green function method~\cite{papanikolaou}. This method is 
ideal for treating systems involving 
impurities on or in surfaces and in bulk crystals. Within the KKR method the 
impurities are described by considering a cluster 
of perturbed atomic potentials which includes the potentials of the impurities 
and the perturbed potentials of several neighbor shells. Also in the vacuum 
region the space is filled by cellular potentials, of which the ones close 
to the impurity are perturbed. The impurity potential and the perturbed 
potentials of the neighboring cells are embedded in 
an otherwise ideal unperturbed surface. 

The KKR method is based on multiple--scattering theory. For non--overlapping 
potentials the following angular momentum representation of the Green's function 
$G({\mathbf r}+{\mathbf R}_n,{\mathbf r'}+{\mathbf R}_{n'};E)$ can be
 derived:  
\begin{eqnarray}
\!\!\!\!{G}({\mathbf r}+{\mathbf R}_n,{\mathbf r'}+{\mathbf R}_{n'};E)\!\!\!&=&\!\!\! 
-i \sqrt{E} \sum_{L}{R}_{L}^n({\mathbf r_<};E){H}_{L}^n({\mathbf r_>};E)\delta_{nn'}\nonumber\\
\!\!\!\!&+& \sum_{LL'}{R}_{L}^n({\mathbf r};E) {G}_{LL'}^{nn'}(E)
{R}_{L'}^{n'}({\mathbf r'};E)
\label{eq:1}
\end{eqnarray}

Here $E$ is the energy and $\mathbf{R}_{n}$, $\mathbf{R}_{n'}$ refers to the 
atomic positions. 
$\mathbf{r_<}$ and $\mathbf{r_>}$ denote respectively the shorter and 
longer of the vectors $\mathbf{r}$ 
and $\mathbf{r'}$ which define the position in the Wigner--Seitz (WS) cell centered around ${\mathbf R}_n$ 
or ${\mathbf R}_n'$. The 
${R}_{L}^n(\mathbf{r};E)$ and ${H}_{L}^n(\mathbf{r};E)$ 
are respectively the regular and irregular solution of the Schr\"odinger 
equation. 

The structural Green functions ${G}_{LL'}^{nn'}(E)$ are 
then obtained by solving the Dyson equation for each spin direction

\begin{eqnarray}
\!\!\!\!{G}_{LL'}^{nn'}(E) &=& \otop{G}_{LL'}^{nn'}(E) \nonumber\\
\!\!\!\!&+& \!\!\!\!\!\!\sum_{n'',L''L'''}\otop{G}_{LL''}^{nn''}(E) 
{\Delta t}_{L''L'''}^{n''}(E) {G}_{L'''L'}^{n''n'}(E)
\label{eq:2}
\end{eqnarray}
The summation in (\ref{eq:2}) is over all lattice sites $n''$ and angular momenta $L''$, 
$L'''$ for which the perturbation 
${\Delta t}_{L''L'''}^{n''}(E)={t}_{L''L'''}^{n''}(E) -\otop{t}_{L''L'''}^{n''}(E) $ 
between the ${t}$ matrices of the real 
and the reference system is significant. $\otop{G}_{LL'}^{nn'}$ are 
the structural Green function of the reference system, {\it i.e.} in our 
case the ideal Cu(111) surface.
 
We have carried out our calculations in the local spin density 
approximation (LSDA) with the parameters of Vosko {\it et al.} \cite{vosko}. Angular 
momenta up to $l_{max} = 3$ are included in the expansion of the wave 
functions and up to $2l_{max} = 6$ in the charge density expansion. We have checked 
these cut-offs to be adequate for our purpose. 

 First, the surface Green functions were determined for the surface of 
Cu(111). The lattice LSDA equilibrium
parameter was used (6.63 {\it a.u.} $\approx$ 3.51 \AA). 
To describe the impurities on the surface, we consider 
a cluster of perturbed potentials which includes the potentials of the impurities 
and the perturbed potentials of several neighbors shell, with 
typical size of 29 perturbed sites for the adatoms. All adatoms are assumed 
to sit at the hollow position in the first vacuum layer.
In the following, we take as reference energy the Fermi level $E_F$.

According to the Tersoff-Hamann model \cite{tersoff} the scanning tunneling 
spectra can be related to the $s$-DOS induced by the surface or by the 
adatom at the position of the STS tip. Adopting this model, we calculate 
the $s$-LDOS at a distance $z$ = 2.86 \AA~directly above the adatom. This corresponds 
to a lattice position in the third vacuum layer above the surface.

\section{Results}

%\subsection{ADATOMS}
\begin{figure}[!ht]
\begin{center}
%\psfigurepath{figures}
\includegraphics*[angle=270,width=\linewidth]{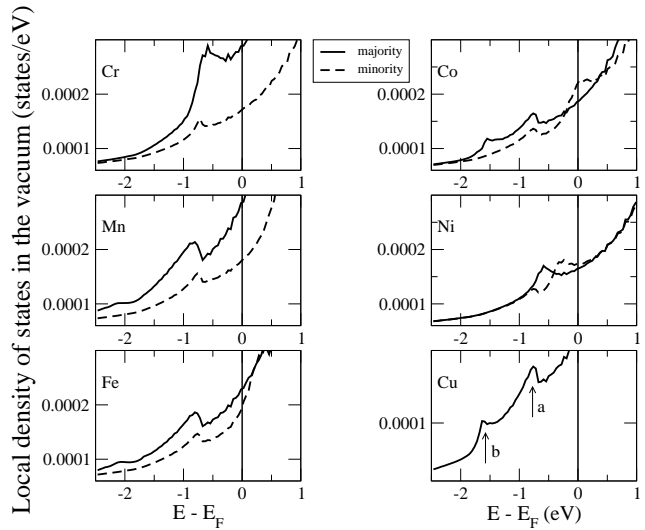}
%\vspace{.14\linewidth}
\caption{Local density of states (LDOS) at the second vacuum layer above the 3{\it d} 
adatoms (2.86 \AA) on the Cu(111) surface. The full lines refer to majority-spin states,
 the dashed lines to minority-spin ones. The two arrows 
show the protrusions discussed in the 
text: (a) corresponds to a split-off state, (b) is a $d_{z^2}$ resonance.}
\label{total}
\end{center}
\end{figure}
The calculated LDOS for 3{\it d} adatoms presented in Fig.\ref{total}
exhibit several adatom--induced peaks. The LDOS refer to an unoccupied 
{\it lattice position} in the third vacuum layer above the surface,
 {\it i.e.} 2.86 \AA~above 
the single adatoms in the first vacuum layer. We focused on 
the region where the split-off state appears experimentally, {\it i.e.}, around
the bottom of the surface state, which in the actual calculations is 
located at $-$0.68 eV. We note that experimentally, the threshold energy is 
higher and is situated at $-$0.45 eV\cite{reinert}. This inconsistency 
is due to the LDA equilibrium lattice parameter we used since a test calculation 
with the experimental lattice parameter gives a value of $-$0.49 eV for the 
threshold energy.

\begin{figure}[!ht]
\begin{center}
%\psfigurepath{figures}
\includegraphics*[angle=270,width=\linewidth]{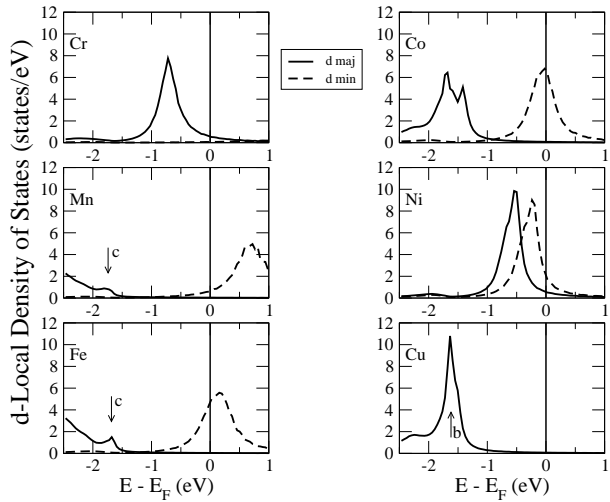}
%\vspace{.14\linewidth}
\caption{{\it d}-contributions to the Local density of states (LDOS) of 3$d$ adatoms 
on Cu(111) surface. The full lines describe the majority-spin states and the dashed 
lines describe the minority-spin states. Since the Cu adatom is non-magnetic, the 
majority and minority virtual bond states coincide.}
\label{total2}
\end{center}
\end{figure}

Let us start with a Cu adatom. Below E$_F$, Fig.\ref{total} shows that 
two states appear in the LDOS. The first one (see arrow (a) in Fig.\ref{total})
 is a split-off state situated at the bottom of the surface state ($\approx$~$-$0.68 eV) 
as was already found by Limot {\it et al.}\cite{limot} and by 
Olsson {\it et al.}\cite{olsson}. To understand the origin of the 
second protrusion (see arrow (b)) we plot in Fig.\ref{total2} 
the $d$-partial LDOS of the adatoms which shows that it comes
 from a resonance of 
the $d_{z^2}$ state at $\approx$~$-$1.7 eV.

For all the 3$d$-adatoms examined, we find for both spin directions a 
peak in the LDOS in the region below the threshold value of the surface state. 
However, we also find peaks at other energies, {\it e.g.} for Cu at 
$-$1.7 eV as mentioned earlier and for Co at $-$1.5 eV for spin-up 
and at around 0 eV for spin-down 
states. In order to understand this behavior, we have in first place to 
understand the electronic and magnetic properties of the impurities and then 
which of the impurity states penetrate well into the vacuum, such that they 
show up in the LDOS of the second layer and can be detected by the STM.

Let us first address the second question. In the vacuum region 
states with small in-plane $\vec{k}$-components ${k}_{||}$ decay most slowly 
as a function of the perpendicular distance z from the surface. In fact, 
for a given energy $E=-\chi^2$ below the vacuum barrier, a wave function 
with in-plane component ${k}_{||}$ decays as 
$e^{-(\chi^2+k^2_{||})^{\frac{1}{2}}z}$. Therefore states with small 
${k}_{||}$-values decay slowest; for higher $k_{||}$ part of the 
kinetic energy is ``absorbed'' by in-plane oscillations. Since 
the states with $k_{||} = 0$ show 
no in-plane oscillations and exhibit the full symmetry of the surface, 
we find that for the (111) surface states with $s$, $p_z$ and $d_{z^2}$ show 
a slow decay in the vacuum region and can be well seen in STM, while 
other $p$- and $d$-states are strongly attenuated. 
An analogous argument holds for single 
adatoms on the (111) surface, since only these states 
exhibit the full point symmetry of the adatom-on-surface 
configuration~\cite{papanikolaou2}, and have thus no oscillations to 
absorb part of the kinetic energy.

To understand the magnetic properties of the adatoms, we have plotted in 
Fig.\ref{total2} the $d$-contribution to the local density of states 
(LDOS) at the adatom site. For the Cu-adatom, we find a sharp $d$-peak at $-$1.7 eV, {\it i.e.} 
at the edge of the bulk $d$-band of Cu. This is a consequence of the 
repulsive potential, which the Cu adatom experiences in the first layer, 
shifting the $d$ states to higher energies than in the bulk. All 
other impurities are magnetic 
and exhibit sizable moments, leading to a spin splitting of the so-called 
virtual bound states. For the different adatoms, the calculated local moments 
M$_s$ are: Cr (4.06 $\mu_B$), Mn (4.28 $\mu_B$), Fe (3.21 $\mu_B$), 
Co (1.96 $\mu_B$) and Ni (0.34 $\mu_B$). Note that the spin splitting is 
roughly given by $I \cdot M_s$, where $I$ is the exchange integral of the order of 
1 eV\cite{gunarsson}. In the 
case of the Cr adatom, the minority peak is at higher energies and cannot 
be seen in Fig.\ref{total2}, while for Fe and Mn the majority peaks are 
at lower energies.

\begin{figure}[!ht]
\begin{center}
%\psfigurepath{figures}
\includegraphics*[angle=270,width=\linewidth]{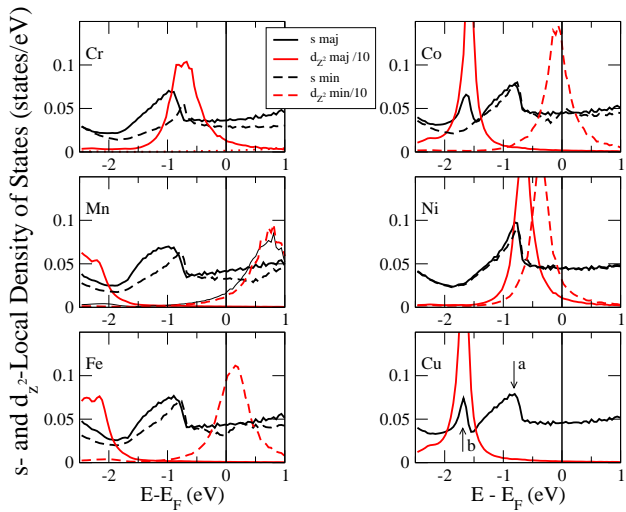}
%\vspace{.14\linewidth}
\caption{(color online) Focus on the $s$-partial LDOS (black curves) of the 
impurity atoms and $d_{z^2}$-partial 
LDOS (red curves) reduced by a factor 10. Full lines represent majority spin, 
dashed lines minority spin.}
\label{total3}
\end{center}
\end{figure}

Let us now come back to the interpretation of Fig.\ref{total}, showing 
the LDOS in the second vacuum layer, at the position above the adatom, for 
majority and minority electrons. Independently of the peak structure, we 
observe a general increase of the DOS at higher energies, 
which arises from the increase of the spatial extent of the wave functions 
for larger energies. For Cu, the peak at $-$1.7 eV coincides with the 
$d_{z^2}$-peak in the local DOS of the adatom, shown in Fig.\ref{total2}. 
For clarification we show in Fig.\ref{total3} the local $s$-DOS and 
the $d_{z^2}$-DOS of the adatoms (the latter reduced by a factor of 
10). For Cu, as well as for the majority states of Co, we see a maximum and 
minimum in the $s$-LDOS at the $d_{z^2}$-peak position, arising from the 
Fano-like resonant scattering of the $s$-states at the $d_{z^2}$-resonance. 
This effect cannot occur for a single adatom in free space or in jellium, 
since the $s$- and $d$-orbitals are orthogonal. Therefore it is brought about 
by the reduced symmetry, {\it i.e.} the scattering at the substrate atoms. 
The LDOS-peak below $-$0.68 eV (arrow (a)) is the split-off state of the Cu adatom, 
induced by the attractive nature of the adatom potential in the first vacuum 
layer. The same states are also seen for the Co-adatom, more or less 
identical for both spin directions. In addition we see for the 
minority $s$-state of Co a Fano-like resonance behavior at the Fermi level, 
arising from the interaction with the minority $d_{z^2}$-virtual bound states.

In the case of the Ni adatoms the virtual bound states for the two spin directions 
are only weakly split and more or less coincide with the energy level of 
the split-off surface state. Therefore in the local $s$-DOS both effects, the 
formation of the split-off state and the resonant scattering at the $d_{z^2}$ 
resonances, cannot be distinguished. However, in the vacuum region (Fig.\ref{total}) 
the spin 
splitting of the majority and minority $d_{z^2}$-states can be clearly seen.

For Fe-, Mn- and Cr-adatoms another effect can be seen in the vacuum LDOS 
(Fig.\ref{total}) and the local $s$-DOS in Fig.\ref{total3}. The intensity 
in the majority split-off surface state is considerably higher than for the 
minority state. This can have several reasons. For instance, due to the 
exchange splitting the majority potential is somewhat stronger than the minority 
one, leading to a smaller lateral extension of the split-off states and to a larger 
intensity on the adatom site. In particular in the case of Cr, also the resonant 
interaction with the impurity $d_{z^2}$ virtual bound state becomes important, 
strongly increasing the majority intensity on the impurity site as well as 
in vacuum.

Moreover, we have noticed a small peak 
appearing in the Mn- and Fe-adatom majority 
$d$-LDOS (see arrow (c) in Fig.\ref{total2}) 
at the same position where the virtual bound states of Cu-adatom is situated 
({\it i.e.} $\approx$ $-$1.7 eV). However these protrusions have a different origin 
since there is no peak at $-$1.7 eV in the $d_{z^2}$-LDOS for Mn- and Fe-adatom 
contrary to Cu- or Co-adatoms (see Fig.\ref{total3}). They appear at the 
remaining $d$-partial LDOS ($d_{xy}$, $d_{yz}$, $d_{xz}$ and $d_{x^2-y^2}$) which 
are strongly damped in the vacuum. For symmetry reasons 
they do not hybridize with the $s$-LDOS explaining thus why we do not 
see a peak at the $s$-LDOS of Mn and Fe-adatoms contrary to Cu and Co-adatoms. 
We believe, however, these peaks can be interpreted as split-off states from a surface 
state at the $\bar M$ point\cite{euceda}, which shows a negative dispersion, such that a repulsive 
impurity potential leads for these $d$-states to a split-off state at higher energies, 
{\it i.e.} above the corresponding surface band.
\begin{figure}[!ht]
\begin{center}
\includegraphics*[angle=270,width=1.\linewidth]{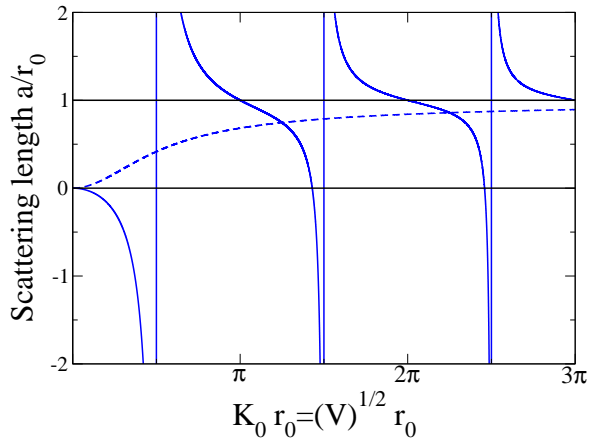}
\caption{(color online) Variation of the scattering length $a$ versus the potential. 
Full lines correspond to a negative potential while blue one describe the case of 
positive potential (always repulsive). The figure is obtained 
with a small positive energy value for an elastic $s$-scattering by a rectangular 
spherical potential depth.}
\label{sinus}
\end{center}
\end{figure}

{\bf {\it sp}-impurities as adatoms:} We consider now some $sp$-impurities 
as adatoms, for which the behavior is not complicated by the spin polarized 
$d$-states. As first candidate we consider Ca at the beginning of the 3$d$-series. 
The calculations give a well defined split-off state at the same position as for Cu 
and the 3$d$ impurities, {\it i.e.} below the minimum of the surface band. 
In addition we perform calculations for Zn, Ga and Ge adatoms. The split-off 
state is still seen for Zn but not anymore for Ga and Ge atoms. The reason for this 
is that in the LDOS of Ga and Ge adatoms the $s$-states have moved to lower 
energies below the surface state minimum. In this case the $s$-scattering 
at the adatoms becomes effectively repulsive, so that no split-off state occurs. 
To explain this we note that the scattering behavior of a scattering center is 
directly related to the $t$-matrix, and only indirectly to the potential. 
Our results are in contradiction with the usual statement that any attractive 
potential leads to a split-off state of a two-dimensional surface state. This is 
not correct in our case, 
since the $t$-matrix of the adatom is basically a three-dimensional quantity.
For $s$-scattering the $t$-matrix is for low energies $E \sim 0$ given 
by the so-called scattering length {\it a}, being discussed in many books 
on quantum mechanics. The quantity {\it a} is the length where the 
extrapolation of the asymptotic form of the wave function for $E = 0$ vanishes. For 
the simple model of spherical potential well of depth $V$ and radius $r_0$ the 
scattering length $a$ is plotted in Fig.\ref{sinus}. For a repulsive potential 
the scattering length is positive and approaches the well radius $r_0$ for 
large $V$. For a weakly attractive potential $a$ is negative. However when the 
potential $V$ becomes stronger attractive, the scattering length assumes more stronger 
negative values, until it jumps. At a critical strength $V = V_0$ from $- \infty$ to 
$+ \infty$, and is positive for further increased $V$-values. At the critical 
strength $V_0$ a bound state appears at $E = 0$, moving to lower energies for 
further increased negative $V$ values, and making the scattering length positive as 
for a repulsive potential. To compare with the real situations of the adatoms, the 
potential of the transition metal atoms is sufficiently weak, that a split-off state 
exist, since the atomic 4$s$-level is far above the Fermi level, and the 
scattering length is negative. However for the Ge adatom the $s$-level has moved below 
the minimum of the surface state, so that the scattering length is positive and 
no split-off state exists. When progressing in the atomic table a split-off 
state can only appear again, when in the next series of the elements, say from 
Rb to Ag, the 5$s$-level has moved down towards the Fermi level.
\begin{figure}[!ht]
\begin{center}
%\psfigurepath{figures}
\includegraphics*[angle=270,width=1.\linewidth]{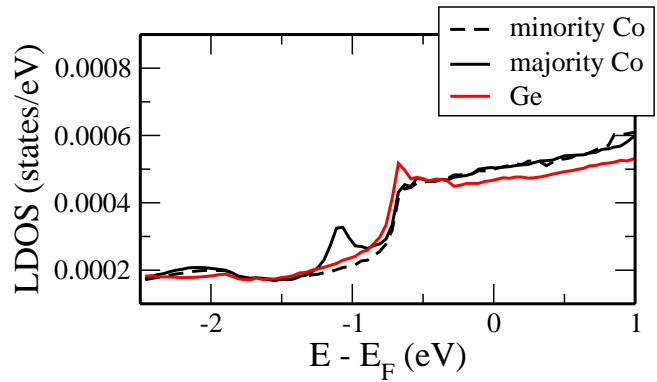}
%\vspace{.14\linewidth}
\caption{(color online) LDOS in vacuum at the second layer above the impurities which 
are sitting in the surface layer. One can see the 
appearance of the split-off 
state above Ge (red line) impurities but not above Co (black line). Above 
the last one, a protrusion 
appears at $\approx$~$-$1.3 eV on the majority spin channel which are 
due to the $d$ state of the Co adatom.}
\label{compare}
\end{center}
\end{figure}

{\bf Impurities in the surface layer:} The scattering of the surface 
states at impurities in the first surface layer is basically different 
from the scattering of $s$-electrons, since the effective potential for scattering 
is the difference between the potential of the impurity and the potential of the 
substituted Cu atom. Therefore all 3$d$-impurities in the first layer do not 
show any split-off surface state, since the potential difference is very small 
(and moreover slightly repulsive). On the other hand for Ge impurities in the first 
layer, the calculations yield a split-off state, as is shown by the small peak 
in Fig.\ref{compare}. Apparently the difference in potential is sufficiently 
attractive, such that a weakly localized state is formed. Therefore we 
obtain the opposite trend as for the adatoms. Transition metal impurities 
exhibit a split-off surface state as adatoms, but not as substitutional impurities 
in the first layer, whereas for Ga and Ge just the opposite is true.

%{\bf Model:} Let us consider the elastic scattering of particles by a central 
%potential V which is non-vanishing in a limited region of space, $\mathbf{r} < d$. 
%Outside the scattering region, the particles move freely. 

%The Schr\"odinger equation reads:
%\begin{eqnarray}
%(\frac{d^2}{dr^2} + k^2 - V(r))R(r)=0,
%\end{eqnarray}
%with the boundary condition:
%\begin{eqnarray}
%R(0)=0
%\end{eqnarray}
%If the potential vanishes the wave function is 
%\begin{eqnarray}
%R_{0}(r)=sin(kr)
%\end{eqnarray}
%Considering a non-vanishing potential leads to a phase shift $\delta$ at large 
%distances:
%\begin{eqnarray}
%R(r)=C sin(kr + \delta)
%\end{eqnarray}
%with C being a constant.

%For small $k$, R(r) behaves like $\approx C k (r-a)$ with $a = -\frac{\delta}{k}$. One 
%notices that outside the scattering region the wave function has no curvature 
%contrary to the inside scattering region.

%{\underline{First case:}} Scattering by a weak attractiv potential well:
%\begin{eqnarray}
%V(r)= \left\{
    %      \begin{array}{ll}
   %         -V_0 & \qquad \mathrm{if}\quad x\leq 0, \quad with \quad V_0 > 0 \\
  %           0 & \qquad \mathrm{if}\quad x > 0, \\
%          \end{array}
 %       \right.
%\end{eqnarray}
\section{Limitations of the LSDA}
We no discuss the limitations of our calculations, in particular 
concerning the Kondo effect which cannot be captured by the LSDA. It is 
well-known that, at low temperatures, the spin moment of the magnetic impurities 
fluctuates, so that these appear non-magnetic. The temperatures at which the experiments 
are conducted are in many cases below the Kondo temperature; {\it e.g.}, a 
characteristic Kondo feature in the spectra was observed for Co on Cu in 
Ref\cite{limot}. 

The Kondo effect is characterized by a narrow Abrikosov-Suhl resonance of the DOS at 
$E_F$ which is absent in our calculations. However, the split-off states are well 
below $E_F$. Furthermore, the Abrikosov-Suhl appears by the interaction of impurity 
$s$-states with the surface band. Therefore our results on the split-off state are 
physically relevant.

On the other hand, the spin-dependent spectra of the magnetic impurities should be 
corrected towards an averaging of the two spin channels, if the temperature is 
below the Kondo temperature. Although the Kondo fluctuations kill the magnetic moment, 
the splitting of the $d$ virtual bound states remains, corresponding to single- 
and double- occupancy of the impurity local state.

\section{summary}
We have performed self-consistent calculations on single impurities 
deposited on Cu(111) surface in order to investigate the 
split-off state recently seen in experiment\cite{limot,olsson}. We show 
the existence of two kinds of 
state localizations: One is due to an attractive potential 
below the $\bar \Gamma$ surface state and the other one is due to a repulsive 
potential above the $\bar M$ surface state. We 
found that Ca, all 3$d$ and Zn adatoms produce the first resonance 
which is not the case for the $sp$ adatoms Ga and Ge even if fundamentally 
it is known that any attractive potential should lead to a 
split-off state at the bottom of a two-dimensional electron state. Its presence for Ca 
means that $s$ states have a stronger contribution to its 
realization then $d$ states. The behavior is totally different if 
the impurities are embedded in the surface layer. In particular, 
Ga and Ge lead to a protrusion at the bottom of the surface state. One 
question addressed in this work is how to explain 
the non-existence of state localizations for $sp$-adatoms and 3$d$-impurities embedded 
in the first surface layer. In the latter case, the reference potential is 
then the Cu one which is not too different from the potentials of all 3$d$ impurities. 
Therefore we do not see any localization. The case of $sp$ adatoms is 
explained by the sign change of the scattering length for attractive potentials, 
which then may act effectively as repulsive potentials. 

\section*{ACKNOWLEDGMENTS}
We would like to thank Laurent Limot and Richard Berndt for 
fruitful discussions. We are grateful to the Deutsche Forschungsgemeinschaft 
for financial support via the Priority Programme ``Clusters in contact with 
surfaces'' (SPP 1153).


\begin{thebibliography}{99}


\bibitem{tamm} {I. E. Tamm, Z. Phys. {\bf 76}, 849, (1932).}

\bibitem{shockley}{W. Shockley, Phys. Rev. {\bf 56}, 317 (1939).}

\bibitem{limot}{L. Limot, E. Pehle, J. Kr\"oger, and R. Berndt, PRL {\bf 94}, 
036805 (2005); J. Kr\"oger, L. Limot, H. Jensen, R. Berndt, S. Crampin and E. Pehlke, 
Progress in Surface Science, {\bf 80},26 (2005).}

\bibitem{olsson}{F. E. Olsson, M. Persson, A. G. Borisov, J. --P. Gauyacq, J. Lagoute 
and S. F\"olsch, P. R. L. {\bf 93}, 206803 (2004).}

%\bibitem{kresse}{G. Kresse and J. Furthm\"uller, Phys. Rev. {\bf B 54}, 11169 (1996).} 

\bibitem{simon}{B. Simon, Ann. Phys. (N.Y.) {\bf 97}, 279 (1976).}

\bibitem{gauyacq}{J. P. Gauyacq, A. G. Borisov, and A. K. Kazansky, Appl. Phys. {\bf 78}, 
                  141 (2004).}

\bibitem{madhavan}{V. Madhavan, W. Chen, T. Jamneala, M. F. Crommie, and N. S. Wingreen, 
Science {\bf 280}, 567 (1998); V. Madhavan, W. Chen, T. Jamneala, M. F. Crommie, and 
Ned S. Wingreen, Phys. Rev. B {\bf 64}, 165412 (2001).}

\bibitem{davis}{A. Davies, J. A. Stroscio, D. T. Pierce, and 
R. J. Celotta, Phys. Rev. Lett. 
{\bf 76}, 4175 (1996).}

\bibitem{papanikolaou2}{N. Papanikolaou, B. Nonas, S. Heinze, R. Zeller, and P. H. Dederichs, 
Phys. Rev. {\bf B 62}, 11118 (2000).}

\bibitem{papanikolaou}{N. Papanikolaou, R. Zeller, and P. H. Dederichs, J. Phys.: Condens. Matter. 
{\bf 14}, 2799 (2002).}

\bibitem{gunarsson}{O. Gunarsson, J. Phys. F: Met. Phys. {\bf 6}, 587 (1976); J. F. 
Janak, PRB {\bf 16}, 255 (1977).}


%\bibitem{SKKR}{R. Zeller, P. H. Dederichs, B. Ujfalussy, L. Szunyogh, and P. Weinberger, Phys. Rev. B {\bf 52}, 8807 (1995);
%R. Zeller, Phys. Rev. B {\bf 55}, 9400 (1997); K. Wildberger, R. Zeller, and P. H. Dederichs, Phys. Rev. B {\bf 55} 10074 (1997).}

\bibitem{vosko} {S. H. Vosko, L. Wilk, and M. Nusair, J. Chem. Phys. {\bf 58}, 1200 (1980).}

\bibitem{tersoff}{J. Tersoff and D. R. Hamann, Phys. Rev. Lett. {\bf 50}, 1998 (1983); 
J. Tersoff and D. R. Hamann, Phys. Rev. B {\bf 31}, 805 (1985).}

\bibitem{reinert}{F. Reinert, G. Nicolay, S. Schmidt, D. Ehrn, and S. H\"ufner, 
Phys. Rev. B {\bf 63}, 115415 (2001).}

\bibitem{euceda}{A. Euceda, D. M. Bylander and D. Kleinman, PRB {\bf 28}, 528 (1983).}
\end{thebibliography}
\end{document}